\documentclass[3p,times]{elsarticle}
\usepackage{ecrc}
\usepackage{amssymb}
\usepackage{siunitx}

\DeclareTextSymbol{\degre}{T1}{6}
\DeclareTextSymbol{\degre}{OT1}{23}

\journalname{Applied Surface Science}
\volume{00}
\firstpage{1}
\runauth{E. B\'{e}villon et al.}

\jnltitlelogo{Applied Surface Science}

\begin{document}

\begin{frontmatter}

%\dochead{European Materials Research Society - Symposium J, laser interaction with advanced materials: fundamentals 
%and applications}

\title{First-principles calculations of heat capacities of ultrafast laser-excited electrons in metals}

\author[ujm]{E. B\'{e}villon}
\author[ujm]{J.P. Colombier\corref{cor}}
\ead{jean.philippe.colombier@univ-st-etienne.fr}
\author[cea]{V. Recoules}
\author[ujm]{R. Stoian}

\cortext[cor]{Corresponding author}

\address[ujm]{Laboratoire Hubert Curien, UMR CNRS 5516, Universit\'{e} de Lyon,\\ Universit\'{e} Jean Monnet 42000,
Saint-Etienne, France}
\address[cea]{CEA-DIF, 91297 Arpajon, France}

\begin{abstract}
Ultrafast laser excitation can induce fast increases of the electronic subsystem temperature. The subsequent electronic 
evolutions in terms of band structure and energy distribution can determine the change of several thermodynamic 
properties, including one essential for energy deposition; the electronic heat capacity. Using density functional 
calculations performed at finite electronic temperatures, the electronic heat capacities dependent on electronic 
temperatures are obtained for a series of metals, including free electron like, transition and noble metals. The effect 
of exchange and correlation functionals and the presence of semicore electrons on electronic heat capacities are first 
evaluated and found to be negligible in most cases. Then, we tested the validity of the free electron approaches, 
varying the number of free electrons per atom. This shows that only simple metals can be correctly fitted with these 
approaches. For transition metals, the presence of localized \textit{d} electrons produces a strong deviation toward 
high energies of the electronic heat capacities, implying that more energy is needed to thermally excite them, compared 
to free $sp$ electrons. This is attributed to collective excitation effects strengthened by a change of the electronic 
screening at high temperature.
\end{abstract}

\CopyrightLine{2014}{Elsevier Ltd. All rights reserved}


\end{frontmatter}

\section{Introduction}

Material response to intense laser excitation is the subject of important research activities, and recent advances have 
revealed the determinant role of primary excitation events. Their accurate comprehension is necessary to correctly 
describe ultrafast structural dynamics \cite{Cavalleri07, Gamaly11}, phase transitions \cite{Povarnitsyn09, 
Recoules06}, nanostruture formation \cite{Colombier12}, ablation dynamics \cite{Lorazo06, Colombier12b}, or strong 
shock propagation \cite{Zhakhovsky11}. The interplay between ultrafast excitation and resulting excited material 
response still requires a comprehensive theoretical description for highly excited solid materials. Ultrashort laser 
irradiation produces a strong inhomogeneous heating of electrons that rapidly exchange energy through electron-electron 
collisions, leading the electronic subsystem to a thermalized state which is intimately related to the electronic 
structure of the material. This fast heating of the electronic subsystem leads in turn to a significant electron-phonon 
nonequilibrium, as the energy of the laser pulse is deposited before relaxation takes on and the material starts 
dissipating energy by thermal or mechanical ways.

During intense laser exposition, the irradiated material undergoes successive stages of extreme constraints, starting 
from the inhomogeneous excitation of electrons, the swift rise of electronic temperature and electronic pressure after 
thermalization of the electronic subsystem on the tens of femtosecond timescale \cite{Mueller13}. These rapid processes 
are followed by the rise of the ionic temperature associated to thermally-triggered phase transitions, before ablating 
or returning to ambient conditions from further energy dissipations occurring at larger timescale. The accurate 
knowledge of these rapidly evolving conditions, in intensity and time, is of importance as they modify material 
properties. Evolution of electronic and ionic temperatures, $T_e$ and $T_i$ respectively, are generally calculated with 
the two-temperature model \cite{Kaganov57,Anisimov74} which involved transient parameters such as electronic heat 
capacity, electron-phonon coupling strength, thermal conductivity, electronic pressure and material absorptivity.

The basic underlying assumption in the two-temperature model is that the electron subsystem consists of a thermalized 
population of a certain amount of electrons, a condition easily fulfilled in ablation processes conditioned by high 
electronic temperatures \cite{Colombier12b}. The energy evolution is determined by photon absorption, energy 
accumulation and thermal/mechanical transport (involving electron-electron and electron-phonon interaction) based on 
nonequilibrium dynamics. The effective number of charge carriers taking part in absorption, storage and dissipation of 
the energy has to be carefully evaluated to give a correct description of the evolution of the corresponding transient 
properties. A certain amount of free electrons responds collectively to the laser field through intraband transitions, 
associated to one-electron excitation corresponding to interband transitions. Modelling this absorption stage is rather 
complex, particularly for ultrashort laser irradiation as it initiates a strong charge disorder. The consequence of a 
nonthermal electron-ion population on the electronic energy storage is also of major importance. Finally, thermal and 
mechanical transports are often described by the concept of "free carriers", since mobility is required for these 
processes. Electronic pressure has been seen to follow a free electron like behavior at high $T_e$, even at solid 
density \cite{Bevillon14}. The electronic heat capacity $C_e$ plays an important role since it connects the quantity of 
absorbed energy to the rise of the electronic temperature of the system, regardless to energy dissipation processes. 
Electronic heat capacities have been derived from post-treatment of first-principles calculations where electronic 
excitation was not fully taken into account \cite{Lin07, Lin08}. Further, these quantities have been refined from 
density functional calculations dependent on the electronic temperature \cite{Bevillon14, Sinko13}.

In this paper, the electronic heat capacity of a series of transition metals is discussed in detail, first according 
to various modelling conditions, considering the effect of the exchange and correlation energy functionals and then the 
effect of semicore electrons. $C_e$ are then discussed in the framework of a free electron approach in order to 
evaluate the impact of $d$ electrons of transition metals on this crucial thermodynamic quantity.

\section{Calculation details}

The modelling of Al, Ni, Cu, Au, Ti and W metals is performed with the code Abinit \cite{Abinit09}, which is based on 
plane-waves description of the wavefunctions. Calculations are carried out within the density functional theory 
\cite{Hohenberg64, Kohn65} extended to finite electronic temperatures \cite{Mermin65}. Projector augmented-waves 
atomic data \cite{Torrent08} are used to take into account the effects of nuclei and core electrons. A cutoff energy of 
40 Ha is applied to restrict the number of plane-waves and the Brillouin zone is meshed with a $30 \times 30 \times 30$ 
$k$-point grid using the Monkhorst-Pack method \cite{Monkhorst76}. The local density approximation (LDA) \cite{LDA-PW92} 
and the generalized gradient approximations (GGA) \cite{GGA-PBE96} are used with or without semicore electrons depending 
on the metals, in order to evaluate effects of semicore electronic states on computed properties. Computations are 
realized at the theoretical equilibrium lattice parameter of the crystal phase obtained at $T_i = 0$ \si{K}. Thus, 
electronic temperatures from $10^{-2}$ to $10^{5}$ \si{K} are applied through a Fermi-Dirac distribution of electrons 
 within a cold lattice. Such high $T_e$ are conjectural and interrogate about phase stabilities. As we need to 
achieve a high temperature asymptotic behavior in this study, we will assume here that time and spatial conditions are 
not fulfilled for phase transitions to occur.

\begin{figure}[ht]
\includegraphics[width=8.4cm]{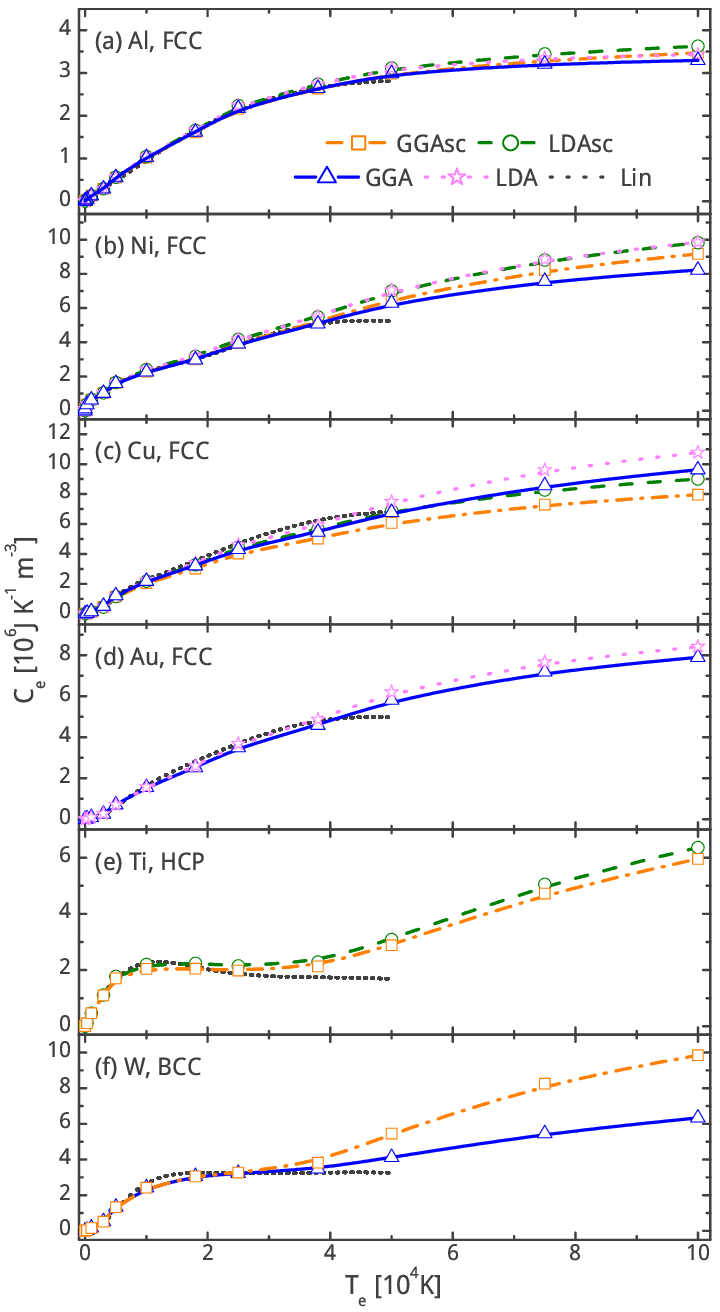}
\caption{\small (Color online) Evolution of the electronic heat capacity with $T_e$ for Al (a), Ni (b), Cu (c), Au (d), 
Ti (e), W (f), using different LDA and GGA exchange and correlation functionals. GGAsc and LDAsc indicate that 
the effect of semicore electrons are taken into account. The dotted black curves represent data of Lin \textit{et al.} 
\cite{Lin08}.}
\label{cepseudo}
\end{figure}

The calculated electronic heat capacity is derived from the variation of the internal energy $E$ with respect to the 
electronic temperature as $C_e = \partial E / \partial T_e$. This thermodynamic quantity is computed for a series of 
metals at various $T_e$ using LDA or GGA functionals with or without semicore electrons. Results are provided in Fig. 
\ref{cepseudo} alongside the values of Lin \textit{et al.} obtained from electronic structures calculated at $T_e = 0 $ 
\si{K} \cite{Lin08}. At a first glance, the agreement between their results and ours is rather good for the low 
temperature range, the deviations appear at $T_e$ above $4 \times 10^{4}$ \si{K}, where electronic structures start to 
react significantly to the heating of the electronic subsystem. In particular, transition and noble metals having a 
$d$-block within their valence band are sensitive to Fermi smearing, impacting $d$ electron population, that induces a 
change of the electronic screening which in turn produces a significant energy shift of the $d$-block \cite{Recoules06, 
Bevillon14}. At high temperature, this produces a significant effect on the band structure, with consequences on the 
electronic heat capacities. The effect of exchange and correlation functional appears to be weak on computed $C_e$. The 
LDA method generally provides lower equilibrium parameters than the GGA, leading to slightly higher electronic 
electronic heat capacities per unit of volume.

The effect of semicore electronic states are also evaluated and Fig. \ref{cepseudo}(f) shows clearly that $4f$ semicore 
electrons of W have an impact at high $T_e$. Despite the fact that the 14 $f$ electrons lie on deep states, around $19$ 
\si{eV} below the valence band, they are significantly impacted by the increase of $T_e$ starting from $4 \times 10^{4}$ 
\si{K}. For Al, Ni and Cu, the highest semicore electronic states correspond to 2$p$ ones, they are located $50$ eV 
below the valence band and are weakly impacted, even at high $T_e$. Nevertheless, a small effect of semicore electronic 
states is noticeable, with stronger $C_e$ values for Al and Ni and lower values for Cu, compared to calculations where 
semicore states are not considered. With first semicore electronic states located at $-27$ \si{eV} for Ti, a significant 
effect of semicore electronic states is likely effective at high $T_e$. A deeper investigation of the evolution of the 
electronic semicore states when $T_e$ increases shows that they are as sensitive as the $d$ band to changes of the 
electronic screening. Depopulation or population of the $d$ band with $T_e$, induces either a decrease or an increase of 
the electronic screening which is signalled by changes of the Hartree energy, and which is effective on localized 
electrons of the system, from $d$ electrons to semicore electrons. As an example, at $10^{5}$ \si{K} the energy of the 
electronic semicore states are shifted for $-3$ \si{eV} for $2p^6$ electrons of Ni, $-8$ eV for $2p^6$ electrons of Cu, 
$2$ \si{eV} for $2p^6$ and $-4$ \si{eV} for $4f^{14}$ electrons of W. Semicore states are not shifted for Al, since this 
metal does not have a $d$ block in its valence band. The shifts of these deep electronic states follow the changes of 
the electronic screening as discussed in Ref. [\citenum{Bevillon14}]. This indicates that the evolution of the 
electronic screening, induced by changes of the electronic population of the $d$ block, propagate to the localized 
charge density corresponding to semicore electronic states.

Nonetheless, due to the low energy of electronic semicore states and their corresponding electronic densities more 
localized around the nucleus, semicore electrons thermalize slower than valence electrons \cite{Fisher06}. According to 
this consideration, the following part of this study relies on PAW atomic data without semicore electrons, at least 
when they are not needed for a good description of general properties. This implies that the GGA functional is used for 
Al, Ni, Cu and the LDA for Au, neglecting the effects of semicore electrons. For Ti and W, the GGA functional is used 
with semicore electrons but without the $4f^{14}$ electrons of W.

\section{Electrons involved in $C_e$}

\begin{figure}[ht]
\includegraphics[width=8.4cm]{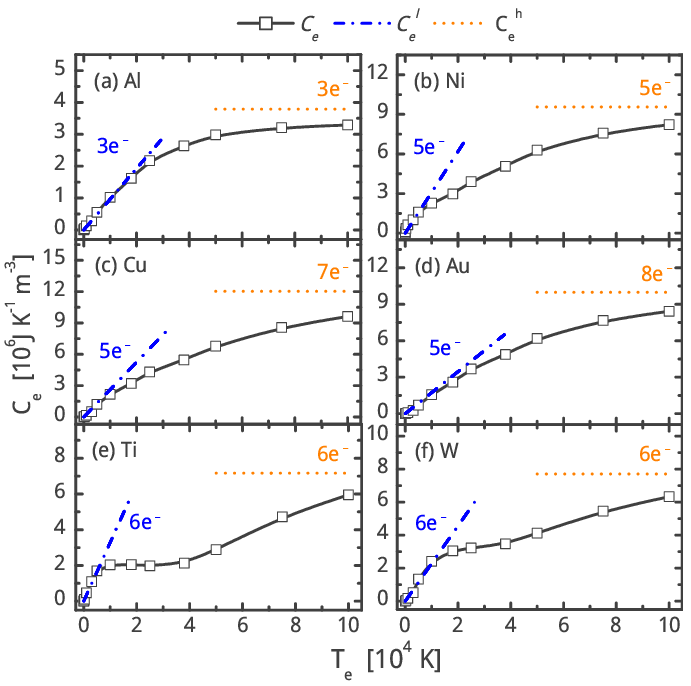}
\caption{\small (Color online) Evolution of the electronic heat capacity with $T_e$ for Al (a), Ni (b), Cu (c), Au (d), 
Ti (e), W (f), and its corresponding low and high temperature limits. The number of electrons used to fit the 
theoretical curve in both models is provided.}
\label{ce}
\end{figure}

In the framework of a free electron model, $C_e$ has a linear behavior at low temperature that can be modelled by a low 
temperature limit $C_e^{l} = \pi^2 N_f n_i k_B^2 T_e/2\varepsilon_F$ and saturates at high temperature to a 
non-degenerate limit $C_e^{h} = 3/2 N_f n_i k_B$. Here, $N_f$ represents the number of free electrons per atom, $n_i$ 
the ionic density, $k_B$ the Boltzmann constant and $\varepsilon_F$ the Fermi level. In Fig. \ref{ce}, $C_e$ is 
plotted for all the studied metals, and free electron like limits are presented at low and high temperatures. These 
limits are fitted to $C_e$ with an effective number of electrons $N_e^{eff}$, provided in the figure, such as $N_f = 
N_e^{eff}$. At low temperature, the fit is performed up to an electronic temperature of $5 \times 10^{3}$ \si{K}, while 
at higher temperature, the corresponding number of electrons is a rough estimation since the asymptotic limit is not 
reached for most of the studied systems.

In order to facilitate the comparison between the number of effective electrons $N_e^{eff}$ required to fit $C_e$ and 
classical free electron numbers, their respective values are reported in Table \ref{nfelt}. $N_f$ from Ref. 
[\citenum{Ashcroft76}] are not designed to evolve with the increase of $T_e$ due to the rigid definition of this number, 
which is based on the most extended orbitals of the electronic configuration of the isolated atoms. On the contrary, 
temperature dependent $N_f(T_e)$ are obtained from extended-orbital overlapping considerations based on 
temperature-dependent electronic structure calculations \cite{Bevillon14}, that allows electronic transfer between bands 
and Fermi-Dirac redistribution of electrons. Unsurprisingly, both free electron numbers remain quite close at low 
temperature, with a significant increase of $N_f(T_e)$ compared to $N_f$ when the temperature increases. This originates 
in the redistribution of electrons from localized $d$ electronic states to delocalized electronic states of higher 
energies.

As expected, both low and high temperature approaches perfectly catch the limit behaviors of the free electron 
metal Al, with an effective number of electrons equals to 3, in agreement with the numbers of free electrons 
expected for this element. However, the number of effective electrons, needed to correctly fit the complex evolution of 
electron heat capacities of transition metals, can be very high and somewhat unrealistic. At low electronic temperature, 
an important gap is observed between $N_e^{eff}$ and expected $N_f$ for transition metals. For W, the number of 
effective electrons equals the number of electrons in the valence band, while for Ti $N_e^{eff}$ is even higher than 
available electrons in the valence band. Reducing the temperature interval where the fit is applied does not modify 
significantly $N_e^{eff}$, except in the case of Ni where this number reaches 19 electrons to catch the particularly 
strong slope of $C_e$ at $T_e = 300$ \si{K} \cite{Lin07}. In all cases, $N_e^{eff}$ remains remarkably high compared to 
the number of free electrons classically used.

\begin{table}[ht]
\begin{center}
\caption{Typical free electron numbers $N_f$ from Ref. [\citenum{Ashcroft76}] and temperature-dependent free 
electrons numbers $N_f(T_e)$ from Ref. [\citenum{Bevillon14}], with estimated number of effective electrons 
$N_e^{eff}$ required to fit $C_{e}$ in the low and high temperature regimes.}
\label{nfelt}
\begin{tabular}{lcccccc}
\hline
                       & Al     & Ni    & Cu   & Au   & Ti   & W    \\
\multicolumn{1}{c}{}   & \multicolumn{6}{c}{$T_e = 5 \times 10^{3}$ \si{K}}  \\
$N_f$                  & 3      & 2     & 1    & 1    & 2    & 2    \\
$N_f(T_e)$             & 3.0    & 1.5   & 1.9  & 2.4  & 1.3  & 2.1  \\
$N_e^{eff}$            & 3      & 5     & 5    & 5    & 6    & 6    \\
\multicolumn{1}{c}{}   & \multicolumn{6}{c}{$T_e = 10^{5}$ \si{K}} \\
$N_f$                  & 3      & 2     & 1    & 1    & 2    & 2    \\
$N_f(T_e)$             & 3.0    & 2.9   & 3.3  & 4.2  & 2.2  & 3.5  \\
$N_e^{eff}$            & 3      & 5     & 7    & 8    & 6    & 6    \\
\hline
\end{tabular}
\end{center}
\end{table}

From the low temperature to the high temperature regime, $N_e^{eff}$ values evolve differently from a material to an 
other, remaining constant for Ni, Ti and W, while increasing for Cu and Au. These values still strongly exceed typical 
values of $N_f$, even if the temperature dependence is considered, signalling a collective effect of $d$-electrons.

\section{Discussion}

\begin{figure}[ht]
\includegraphics[width=8.4cm]{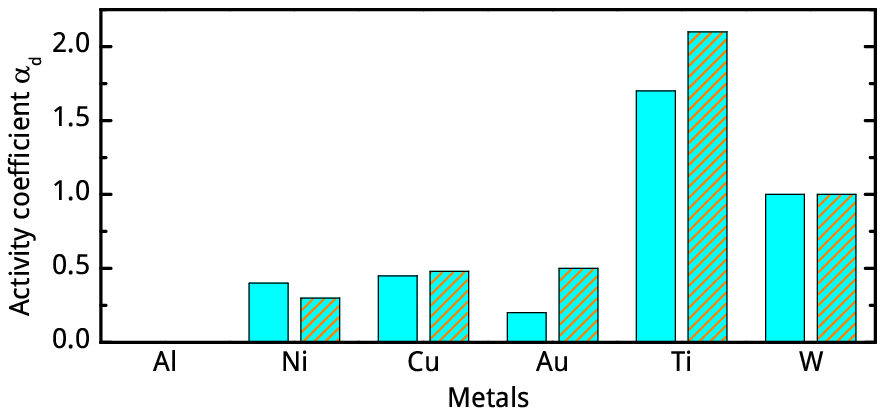}
\caption{\small (Color online) Histogram representation of activity coefficients of localized $d$ electrons. For each 
material, the left histogram corresponds to the low temperature case while the right one (patterned) represents the high 
temperature case.}
\label{alpha}
\end{figure}

The significant differences between $N_e^{eff}$ and the typical values of $N_f$ at low and high temperatures, 
demonstrate the inability of asymptotic limits derived from free electron approaches to grasp the complex behavior of 
transition metals, which relies on the presence of localized $d$ electrons. Indeed, these localized electrons also 
contribute to the change of the electronic heat capacity, through the change of free energy in the system as $T_e$ 
increases. The electronic heat capacity informs about the amount of energy needed to rise the electronic temperature of 
a given system. The fast deviation toward high values of $C_e$ observed for transition metals, which is illustrated by 
values of $N_e^{eff}$ much higher than the classical values of $N_f$, implies that the energy required to heat the 
localized $d$ electrons is higher than the energy needed to heat free electrons. Moreover, non-linear evolution of 
$N_e^{eff}$ with the increase of $T_e$ also indicates a complex impact of $d$ electron contribution to $C_e$. Finally, 
previous studies showed that the change of $d$ electron numbers with $T_e$ modifies the electronic screening that in 
turn induces an energy shift of the $d$-block affecting all localized $d$ electrons \cite{Recoules06,Bevillon14}. 
Consequently, the effect of $d$ electrons on $C_e$ has to be considered as a collective effect of the overall localized 
\textit{d} electrons. Thus, the number of electrons to be considered in the electronic heat capacity limits at low and 
high temperature should be reevaluated, according to an activity coefficient applied to the different types of 
electrons:

\begin{eqnarray}
   \displaystyle N_e^{eff} = \alpha_{f} N_f + \alpha_{d} N_d + \alpha_{sc} N_{sc}
\end{eqnarray}

\noindent where $N_f$ represents the number of free electrons as defined before, $N_d = N_v - N_f$ is the number of 
$d$ electrons, with $N_v$ being the number of valence electrons, and $N_{sc}$ is the number of semicore electrons. 
$\alpha_i$ stands for the activity coefficient for each category of electrons. $\alpha_{f}$ is set to the value of $1$ 
for all metals as it corresponds to the free electron approaches. Since PAW atomic data used here exclude most of the 
semicore electrons and $N_{sc}$ is set to $0$. Considering that $N_e^{eff}$ is equal to the values provided in Fig. 
\ref{ce} it is then possible to deduce the activity coefficients of $d$ electrons for all the transition metals at low 
and high temperatures from $\alpha_d = (N_e^{eff} - N_f) / N_d$. This is solved within a temperature dependence of free 
electron numbers and $d$ electrons numbers that are presented in Table \ref{number-elec}. The corresponding activity 
coefficients of $d$ electrons are provided in Fig. \ref{alpha}.

\begin{table}[ht]
\begin{center}
\caption{Temperature-dependent free electron numbers $N_f(T_e)$ from Ref. [\citenum{Bevillon14}] and corresponding 
number of localized \textit{d} electrons $N_d(T_e) = N_v - N_f(T_e)$ in the low and high temperature regime.}
\label{number-elec}
\begin{tabular}{lcccccc}
\hline
    & \multicolumn{2}{l}{$T_e = 5 \times 10^{3}$ \si{K}}   & \multicolumn{2}{l}{$T_e = 10^{5}$ \si{K}}  \\
    & $N_f(T_e)$ & $N_d(T_e)$ & $N_f(T_e)$ & $N_d(T_e)$  \\
Al  &  3.0  &  0.0  &  3.0  &  0.0  \\
Ni  &  1.5  &  8.5  &  2.9  &  7.1  \\
Cu  &  1.9  &  9.1  &  3.3  &  7.7  \\
Au  &  2.4  &  8.6  &  4.2  &  6.8  \\
Ti  &  1.3  &  2.7  &  2.2  &  1.8  \\
W   &  2.1  &  3.9  &  3.5  &  2.5  \\ 
\hline
\end{tabular}
\end{center}
\end{table}

As expected, $\alpha_{d}$ for Al remains equal to $0$ due to the absence of $d$ electrons. For Ni, Cu and Au, the 
coefficient values are in the range of 0.2-0.5, with an increase for Cu and Au and a decrease for Ni when $T_e$ 
increases. These significant activity coefficients of $d$ electrons - that applies on all localized $d$ electrons 
- imply that the response of these electrons is strong, even when the Fermi smearing weakly affects them at low 
temperature, as for Cu and Au. On the contrary, $\alpha_{d}$ reach high values for Ti and W, around $2$ and $1$ 
respectively and slowly evolve with $T_e$. The strong values obtained for Ti and W originate from the fact that the 
Fermi level lies within the $d$ block, even when $T_e$ increases. Excitation effects directly involves population or 
depopulation of localized $d$ electronic states with strong impact on $C_e$. With a Fermi energy located on the top of 
the $d$ block at low electronic temperature, a similar effect should also be observed for Ni, but this is actually 
hidden by the fit on large range of electronic temperatures. As we will see later, this phenomenon becomes visible when 
the fit is performed at lower temperatures. For W, once $\alpha_{d}$ is obtained from calculations that neglect the 
effect of semicore electrons, it is then possible to use this value to derive $\alpha_{sc}$ from the $C_e$ curve 
obtained by taking into account the $4f^{14}$ electrons (GGAsc curve in Fig. \ref{cepseudo}(f)). This leads to a new fit 
in the high temperature limit, with  $N_e^{eff}$ being equal to $9$. Accordingly, $\alpha_{sc}$ equals to $0$ at low 
temperature and reaches the value of $0.2$ at high temperature, confirming the effect of these electrons on $C_e$, 
assuming they are thermalized.

Since $C_e$ is supposed to evolve linearly in the low temperature regime, it is often characterized by the electronic 
heat capacity coefficient $\gamma = C_e/T_e$. Values at $5 \times 10^{3}$ \si{K} are provided in Table 
\ref{free-electrons-5000k}, where a relative good agreement is found between calculated values and the low temperature 
free electron limit $\gamma^{l}$ based on $N_{e}^{eff}$. However, for Ni the linear evolution of $C_e$ is not respected, 
and a strong change of the slope occurs at low temperature. At $T_e$ equals to $300$ and $10^{3}$ \si{K}, the 
corresponding $\gamma$ values reach $1183.3$ and $631.1$ \si{J K^{-2} m^{-3}} respectively. In the low temperature 
limit, this involves very high values of $N_{e}^{eff}$, $19$ and $10$ respectively, leading to activity coefficients of 
$d$ electrons of $2.1$ and $1.0$ respectively. This strong change of $C_e$ slope at low temperature corresponds to a 
major difference in the behavior of Ni compared to Cu and Au. This difference of behavior originates from the position 
of the Fermi level, at the top of the $d$ block for Ni while it is higher for Cu and Au, due to difference in their 
number of valence electrons. Thus, at low temperature, depopulation of the $d$ block is direct for Ni, while the $d$ 
block is not impacted yet for Cu and Au. When $T_e$ increases, the Fermi level shift toward energies higher than the $d$ 
block for both elements, inducing a strong changes of $\alpha_{d}$ in case of Ni, that progressively tends toward the 
values of Cu and Au. As in the case of Ti and W, a direct impact of the Fermi smearing on the $d$ block has a strong 
effect on $C_e$ which is signalled by important activity coefficient of localized $d$ electrons. To a lower degree, a 
similar deviation is observed for Ti around $T_e =10^{3}$ \si{K}, with a $\gamma$ coefficient reaching the value of 
$466.3$ \si{J K^{-2} m^{-3}}. This is fitted by a value of $N_{e}^{eff}$ equals to $8$ and leads to an even higher 
$\alpha_{d}$ with the value of $2.5$. For these elements, the non-linear evolution of $C_e$ at low electronic 
temperatures leads to strong modifications of subsequent derived quantities. Thus, beyond these general observations, 
the evolution of these activity coefficients remains difficult to be evaluated accurately and values cannot be 
extrapolated to other elements or temperature regime.

\begin{table}[h]
\begin{center}
\caption{Coefficient of electronic heat capacity ($\gamma = C_e/T_e$, \si{J K^{-2} m^{-3}}), at $5 \times 10^{3}$ 
\si{K}.}
\label{free-electrons-5000k}
\begin{tabular}{lcccccc}
\hline
                   &   Al    &   Ni     &   Cu    &    Au    &   Ti    &   W     \\
 $\gamma$          &  110.4  &  320.0   &  243.9  &  141.1   &  340.9  &  266.8  \\
% $\gamma^{exp}$   &  135.0  &  1077.4  &   96.8  &   67.6   &  328.9  &  137.3  \\
 $\gamma^{l}$      &   96.2  &  309.3   &  262.5  &  172.8   &  328.7  &  342.2  \\
\hline
\end{tabular}
\end{center}
\end{table}

\section{Conclusion}

In this paper, the electronic heat capacities are obtained from first-principles calculations performed at finite 
electronic temperatures. The effect of exchange and correlation functional on this thermodynamic quantity is first 
discussed. At high electronic temperature a small difference is observed between LDA and GGA functionals that was 
attributed to different equilibrium lattice parameters. The impact of semicore electrons, is also tested and 
demonstrated at high temperature for 4$f^{14}$ electrons of tungsten. This is dependent on the energy depth of the 
corresponding semicore electronic states and is mostly negligible even at high temperatures, except for W. Assuming 
negligible effects or lower thermalization times for these electrons, effect of semicore electrons were neglected in the 
present study. Electronic heat capacities are also compared to previous theoretical predictions obtained from electronic 
structures not relaxed with the electronic temperature, with a very good agreement up to $4 \times 10^{4}$ \si{K}. 
Beyond this temperature, the response of the electronic structure exhibits differences.

The obtained electronic heat capacities are then discussed in the framework of low and high temperature limits of a 
free electron approach. Both are based on a free electron number, which is modified up to an effective number of 
electrons in order to fit the theoretical values of $C_e$. This approach applies correctly to Al as a free electron 
metal, but a significant overestimation of free electron numbers is found for the transition metals. Assuming a 
collective contribution of localized electrons on the $C_e$ evolution, $d$ electrons where considered associated to an 
activity coefficient. The high values obtained for this coefficient tend to corroborate the fact that $d$ electrons 
respond collectively to the electronic excitation. A strong effect is observed from the impact of Fermi smearing on 
$d$ electrons, as showed for Ni at low temperature and Ti and W in the whole range of temperature. For W, at very high 
temperature, an important contribution of semicore electrons $4f$ is also deduced from this approach, assuming that they 
are thermalized. At high temperature, the change of the electronic screening is found to affect significantly the 
localized $d$ electrons, with a global shift of the $d$-block, strengthening the collective behavior of these electrons.

Additionally, the non-linear effect of $d$ electrons, which is shown by an important dispersion of their activity 
coefficient, illustrates their non-free behavior and their interactive nature. The large numbers of electrons needed to 
fit $C_e$ evolution in the low and high temperature regime is a signature of localized electron influence, mostly $d$ 
electrons but also semicore electrons at very high temperature. Consequently, an unique free electron number cannot 
encompass the complex evolution of the electronic heat capacity and localized electrons, especially $d$ ones, have to be 
considered. This shows that the free electron number is only one of the components on the energy storage of an 
irradiated material, and that localized electrons contribute as well, to various degrees depending on the electronic 
temperature.

\section{Acknowledgments}

 This work was supported by the ANR project DYLIPSS (ANR-12-IS04-0002-01) and by the LABEX MANUTECH-SISE 
(ANR-10-LABX-0075) of the Universit\'{e} de Lyon, within the program "Investissements d'Avenir" (ANR-11-IDEX-0007) 
operated by the French National Research Agency (ANR). Numerical calculations have been performed using resources from 
GENCI, project gen7041. 

\bibliographystyle{elsarticle-num}

\end{document}